\documentclass[aps,pre,twocolumn,showpacs,preprintnumbers,superscriptaddress]{revtex4}
\usepackage{epsfig}

\begin{document}                

\title{Memory difference control of unknown unstable fixed points:
\\Drifting parameter conditions and delayed measurement}
\author{Jens Christian Claussen$^1$,
Thorsten Mausbach$^2$, Alexander Piel$^2$,
 and Heinz Georg Schuster$^1$ }
\address{$^1$Institut f\"ur Theoretische Physik und Astrophysik,
Christian-Albrechts-Universit\"at, 24098 Kiel, Germany
\\ $^2$Institut f\"ur Experimentelle und Angewandte Physik, 
Christian-Albrechts-Universit\"at, 24098 Kiel, Germany \\[3mm] }
\date{July 8, 1998. Published in: \underline{Phys. Rev. E {\bf 58}, 7256-7260 (1998)}}

\begin{abstract}                
Difference control schemes for controlling unstable
fixed points become important if  
the exact position of the fixed point 
is unavailable or moving due to drifting parameters.
We propose a memory difference control method for stabilization of
{\sl a priori} unknown unstable fixed points by 
introducing a memory term.
If the amplitude of the control 
applied in the previous time step
is added to the present control signal,
fixed points with arbitrary
Ljapunov numbers can be controlled.
This method is also extended to compensate
arbitrary time steps of measurement delay.
We show that our method stabilizes orbits of the
Chua circuit where ordinary difference control fails.
\end{abstract}


\maketitle

\section{Introduction}

Two main classes of control strategies are established
in chaos control:
The OGY control algorithm~\cite{ogy90},
almost a standard method for controlling chaos,
does not provide any readjustment
of the fixed point during in-situ measurements without loss of control.
In many experimental systems, however, 
it is desirable to use a control strategy that
does not rely on the knowledge of the exact position of the fixed point,
because the location of the fixed points 
can change due to drifts in parameters.
On the other hand, the time-continuous control method proposed by
Pyragas~\cite{pyragas92} is practically limited by the required 
sampling rate, and does not allow stabilization of
arbitrary orbits as has recently been shown in~\cite{just}.

Both problems are circumvented by 
simple time-discrete difference control~\cite{Bielawski93a}.
It is limited to a certain range of
Ljapunov numbers. Control of arbitrary periodic orbits 
can be achieved if the algorithm is applied only every second 
period~\cite{Bielawski93a,schusterstemmler}
or by a memory method.

In this paper we present an improved memory difference control (MDC)
method that takes 
control amplitudes into account that were
applied at previous time steps.
MDC allows one to stabilize drifting 
fixed points with arbitrary Ljapunov numbers 
and shows an enlarged region of stability.

This method is generalized when dealing with
measurements delayed by $\tau$ time steps (orbit periods).
This task is accomplished
by increasing the number of memorized control amplitudes by~$\tau$.
Given the stable and unstable directions 
of the fixed point with sufficient
accuracy, only one accessible control parameter 
for each unstable direction is required to achieve control.

We compare difference control
 and MDC at the well-known Chua oscillator~\cite{chua} and
 show that orbits for which difference control fails
are stabilized by MDC.

\section{Stabilization of fixed points 
by difference control}
In experimental situations a Poincar\'e section is commonly used
to reduce the dynamics to a time-discrete description by an iterated map
\begin{eqnarray}
\vec{x}_{t+1}= \vec{f}(\vec{x}_t,\vec{r}). \label{dynamik_1dim_full}
\end{eqnarray}
Here $\vec{r}$ denotes a set of control parameters that are
in the unperturbed dynamics
assumed to be constant or varying on a slow time scale compared to the 
the length of a period orbit.

The idea to control chaos by small perturbations of control
parameters implies that  $\vec{r}$ becomes time-dependent.
The time-dependent control parameter 
$\vec{r}_t$ is updated at each discrete time step 
defined by the Poincar\'e section.
Its value is determined according to the specific control algorithm 
and is kept constant for a part of the orbit.
Without loss of generality
we choose $\vec{x}^*=\vec{0}$ and  $\vec{r}=\vec{0}$ for the fixed
point to be stabilized.
The linearized equation of motion around the unstable fixed point
then becomes
\begin{eqnarray}
\vec{x}_{t+1}= L \vec{x}_t + M \vec{r}_t. \label{dynamik_1dim}
\end{eqnarray}
where
\begin{eqnarray}
L_{ij} := \left. \frac{\partial {f}_i}{\partial {x}_j}
\right|_{\vec{x}^*=0, \vec{r}=0}
~~~~ {\rm and} 
~~~~
M_{ij} := \left.   \frac{\partial {f}_i}{\partial {r}_j}
\right|_{\vec{x}^*=0, \vec{r}=0}.
\end{eqnarray}
The Ljapunov numbers of the orbit are the eigenvalues of $L$.
Here one has to distinguish the Ljapunov number of an orbit (or
time-discrete map) from the local (or conditional) Ljapunov exponent 
and the commonly used global Ljapunov exponent
being an ergodic average over the attractor~\cite{schusterbuch}.
In principle it is possible to proceed with a multiparameter 
control by using as many control parameters
as there are degrees of freedom, i.~e.
${\rm{}dim}(\vec{r})={\rm{}dim}(\vec{x})$. 
Instead, it is common to follow Ott, Grebogi and Yorke~\cite{ogy90}
to transform the system to the eigensystem of $L$. 
Control is applied
{\em only} in the unstable subspace~\cite{dynamicsreduce}.
The evolution of the equation of motion
is again of the form of eq. (\ref{dynamik_1dim}) with reduced dimension 
of $L$.

\begin{figure}[thbp]  
\epsfig{file=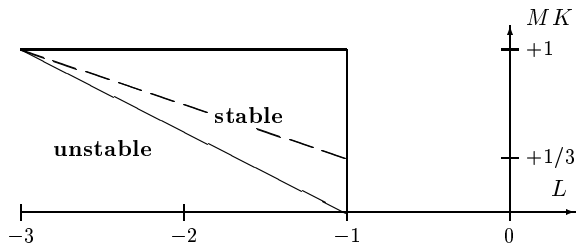, angle=0,width=3in}
\caption{
Stability area for difference control
(one unstable Ljapunov number $L$):
For $|L|<1$ the system is stable without control.
Fixed points with
Ljapunov number
$-3<L<-1$
 can be controlled if $M\cdot{}K$
is chosen within the area bounded by the triangle.
The line within the triangle shows the optimal value for $M\cdot{}K$.
\label{fig:cdcstab}}
\end{figure}

In difference control \cite{Bielawski93a}
the control parameter is updated at the Poincar\'e section
according to 
\begin{eqnarray}
\vec{r}_t = K (\vec{x}_{t}-\vec{x}_{t-1}).
\end{eqnarray}
In contrast to OGY control, difference control 
is limited to fixed points 
with Ljapunov numbers between $-3$ and $-1$
\cite{Bielawski93a,schusterstemmler}.
A stability diagram 
\cite{eigvalueseq6,schusterstemmler} 
for the case of one unstable eigenvalue
is shown in Fig.~\ref{fig:cdcstab}.

\typeout{Fig.1 goes here!}

Simple difference control uses information that is partially 
out of date, resulting in an additional degree of freedom from the 
delayed amplitude $\vec{x}_{t+1}$.
This fact is illustrated by imagining two agents controlling 
the same system in turns. If they do not communicate, control fails 
because of the inherent delay of the system to be controlled.
This effect is compensated when using the information contained 
in the previous value of the control amplitude $\vec{r}_{t-1}$.

\section{Memory difference control} \noindent
We define memory difference control by~\cite{remarkmemoryinx}
\begin{eqnarray}
\vec{r}_t = K (\vec{x}_t-\vec{x}_{t-1}) + N \vec{r}_{t-1}.
\end{eqnarray}
Combined with (\ref{dynamik_1dim}), we obtain a dynamical system 
which reads in delayed coordinates for $\vec{x}$ and $\vec{r}$
\begin{eqnarray}
\left(\begin{array}{c}\vec{x}_{t+1}\\\vec{x}_{t}\\\vec{r}_{t}
\end{array}\right)
=\left(\begin{array}{ccc}L+MK&-MK&MN\\1&0&0\\K&-K&N\end{array}\right)
\left(\begin{array}{c}\vec{x}_{t}\\\vec{x}_{t-1}\\\vec{r}_{t-1}
\end{array}\right)
\label{eq:mdcdynamik}
\end{eqnarray}
In order to stabilize the fixed point all eigenvalues $\alpha_i$ of the 
matrix in (\ref{eq:mdcdynamik}) must have a modulus smaller than one.
This ensures exponential convergence to the fixed point.
If all parameter values are chosen such that
all eigenvalues become zero, the fixed point is reached
after a finite number of time steps.
In fact this can be guaranteed by MDC.
We first assume that $M$ and $(L-1)$
are both invertible, and that 
the number of accessible control parameters 
is equal to or greater than the dimension of the unstable manifold,
i.~e. $\dim(\vec{r})\geq\dim(\vec{x})$.
Then all eigenvalues are zero  \cite{claussenthesis,thirditerate} if
\begin{eqnarray}
\nonumber
K &=& - M^{-1} L^2 (L-1)^{-1} \\
N&=&M^{-1} L (L-1)^{-1} M.
\label{eq:verbdiffkont}
\end{eqnarray}
\typeout{HIER NUMERIERUNG A/B?}

The concept of MDC
can be generalized to stabilization of (known and unknown)
fixed points when $\vec{x}$ can be measured only after a finite number
of delay steps\cite{claussenthesis,claussen98}:
If the system is measured with $\tau$ steps delay, 
(\ref{eq:verbdiffkont}) is replaced by
\begin{eqnarray}
K   &=& -M^{-1} L^{\tau+2} (L-1)^{-1}  \\
\forall_{1\leq{}i\leq{}\tau} ~~~~~~~~
N_i &=& -M^{-1} L^{i} M 
\\ N_{\tau+1} &=& M^{-1}  L^{\tau+1} (L-1)^{-1} M 
\end{eqnarray}
where the feedback now contains a sum over $\tau+1$ preceding
control amplitudes:
\\
\begin{eqnarray}
\vec{r}_t= K(\vec{x}_{t-\tau}-\vec{x}_{t-\tau-1}) 
+\sum_{i=1}^{\tau+1} N_{i} \vec{x}_{t-\tau-i} 
\end{eqnarray}
A similar control scheme can be applied for OGY control
by choosing $K= -M^{-1} L^{\tau+1}$ and $N_i =-M^{-1} L^{i} M$,
 \mbox{($1\leq{}i\leq{}\tau$)}.
For details see \cite{claussenthesis,claussen98}.

\section{The stability diagram in the one-dimensional case}
\begin{figure}[htbp]
\epsfig{file=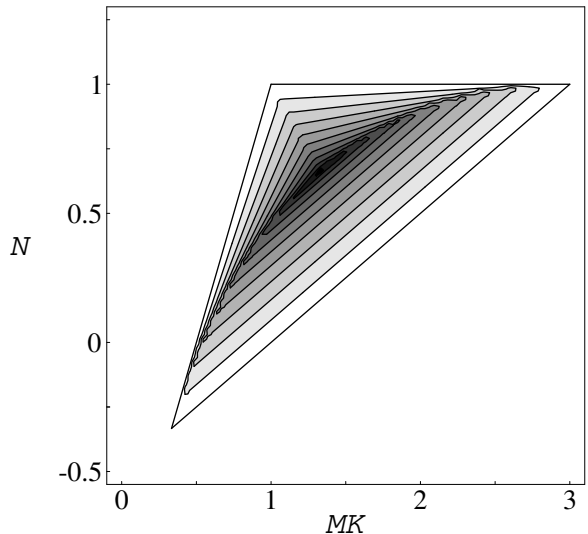, angle=0,width=3in}
\caption{
Stability region of memory difference control for $L=-2$
in the $(MK,N)$-plane.
Within the triangle MDC is stable; the contour lines show
the decrease of the largest eigenvalue  to zero
 in $(MK,N)=(4/3,2/3)$.
\label{fig:diffkontrdreieck}}
\end{figure}

Since the optimal control values are never exactly matched in experiments,
it is important
to know the complete stability diagram in $(K,N)$, 
in particular the optimal value of $K$ for given
$N$ and vice versa.
In the one-dimensional case, the characteristic equation is given by
\begin{eqnarray}
0 = \alpha [\alpha^2 - (L+MK+N) \alpha + (MK+N L)].
\end{eqnarray}
The stability region in the $(K,N)$-plane,
i.e. where all eigenvalues have modulus
smaller than one,
is the triangle 
shown in Fig.~\ref{fig:diffkontrdreieck}.
Its corners are given by 
\begin{eqnarray}
(MK,N)_{+1,+1} &= &(1-L,1) \\
(MK,N)_{-1,-1} &= &\bigg(-\frac{(L+1)^2}{(L-1)},\frac{(L+3)}{(L-1)}\bigg)\\
(MK,N)_{+1,-1} &= &(-1-L,1)\\
\nonumber
\end{eqnarray}
where the eigenvalues take the values $+1$ and $-1$ as 
indicated by the indices.
Two sides of the triangle are determined by the conditions that one 
eigenvalue is equal to $+1$ and $-1$, respectively.
The third is given by $MK+NL=+1$ where the eigenvalues are a complex
conjugate pair on the unit circle.
The line that determines 
that $N$ with
minimal eigenvalues
for a given $K$ (and vice versa)
is given by the algebraic expression
\begin{eqnarray}
0=(MK)^2 + 2MKN+N^2 + (2L-4) MK - 2LN + L^2 \nonumber
\end{eqnarray}
where both eigenvalues coincide.

\typeout{Fig.2 goes here!}

\section{The Chua oscillator \label{chuasection} }
We demonstrate the efficiency of the improved difference method
by controlling unstable periodic orbits of the chaotic attractor of 
Chua's oscillator \cite{chua}. 
The Chua circuit is an autonomous system.
The Poincar\'e section 
necessary for control
is obtained by 
an electronic zero-crossing-detector.  
The frequency is in the range of
$\nu={2\pi}/{\sqrt{LC_1}}\sim 3.6{\rm{}kHz}$ 
and allows the usage of 
digital signal processing tools to implement control algorithms.

Parameter drifts, e.g.~temperature drifts, naturally occur in 
electronic circuits and 
difference control methods have the advantage that they follow the 
drifting fixed point.
In order to get access to an appropriate control
parameter, a VCR (voltage controlled resistor) has been added to the
circuit. The basic dynamics is nevertheless determined by the 
resistor $R$. 

Furthermore the Chua circuit allows to investigate
interesting ranges of Ljapunov numbers simply by choosing different values of 
the main control parameter $R$.

\typeout{Fig.3 goes here!}

\begin{figure}[htbp]
\epsfig{file=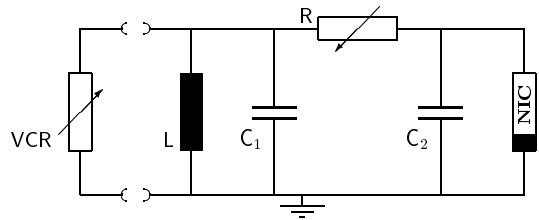, angle=0,width=3in}
\caption{
The Chua circuit: The negative resistance (NIC) is both
nonlinearity and energy source of the circuit. Rough adjustment of
the control parameter can be done by adjusting {\sf R}.
Control is applied with the voltage-dependent resistor (VCR).
\label{fig:circuit}}
\end{figure}

The dynamics of the Chua circuit 
is described, in first approximation,
by the normalized equations
\begin{eqnarray}
\nonumber \dot{u} &=& \alpha (v-u-f(u)) \\
\dot{v} &=& u-v+w \\
\nonumber \dot{w} &=& -\beta v  
\end{eqnarray}
where $f$ is the input-output function of the negative resistance 
approximately described by the piecewise linear descending function 
\begin{eqnarray}
f(u)=m_0u+\frac{1}{2}(m_1-m_0)(|u+1|-|u-1|)
\end{eqnarray}
with $m_0>m_1$ \cite{kapitaniak}.
Rather than solving these equations numerically, we demonstrate 
the stabilization of an unstable periodic orbit directly in the
electronic system.

\typeout{Fig.4 goes here!}

\begin{figure}[htbp]
\epsfig{file=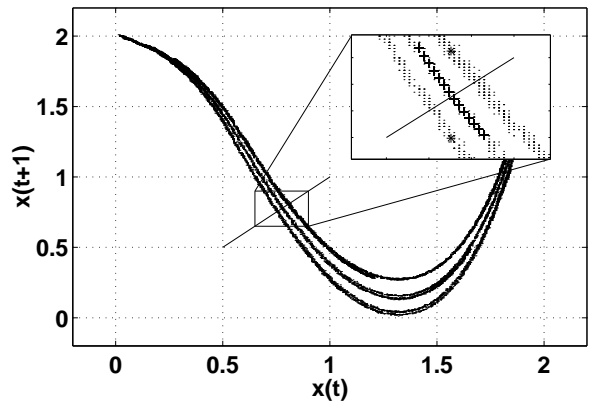, angle=0,width=3in}
\caption{
Return map $x_{t+1}(x_t)$ in the
Poincar\'e{} surface of section 
for three different values of the control parameter.
From this data one obtains values of $L$ and $M$ to adjust the
optimal parameters of control.
\label{poincarefit}}
\end{figure}

\section{Improved Difference Control of the Chua system}
 \label{chuacontrolsection} 

\begin{figure}[htbp]
\epsfig{file=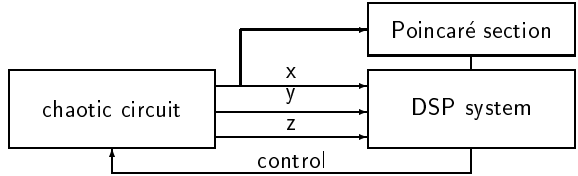, angle=0,width=3in}
\caption{Implementation of the control method:
One coordinate is used to generate the Poincar\'e section.
A digital signal processor is used for measurement, computation, and
application of the control amplitude. 
\label{fig:scheme.eps}}
\end{figure}

The standard control strategy
is to measure the required system
variables, generate the Poincar\'e map 
for e.~g. three adjacent  values of the control parameter
(Fig.~\ref{poincarefit}), and to calculate the parameters of the feedback
to the control parameter (Fig.~\ref{fig:scheme.eps}). 

In the present case the return map $x_{t+1}(x_t,z_t)$ is approximately a 
function of $x_t$ alone.
Therefore only two variables are required. 
The first one, $y_t$, is used to determine 
the Poincar\'e surface of section by a zero crossing detector.
The second one, $x_t$, is used for the calculation of the control.

\typeout{Fig.5 goes here!}

\begin{figure}[htbp]
\epsfig{file=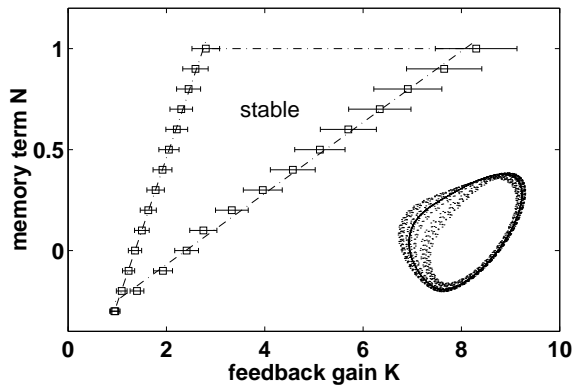, angle=0,width=3in}
\caption{Stabilization of an orbit in the single-scroll chaotic attractor:
Within the measured triangle memory difference control stabilizes the 
orbit of the Chua circuit.
The special case of simple difference control is given by $N=0$.
The inset shows the attractor
and the stabilized periodic orbit.
\label{knp1sc}}
\end{figure}

The stability region for different values of the memory
term $N$ and feedback gain $K$ is measured by changing
the values until control is entirely lost. 
The lower bound of $K$ is easily recognized by a doubling of the 
stabilized period.
However, the upper bound of $K$, where the loss of control is
noise-induced, is more difficult to estimate. 
In Fig.~\ref{knp1sc} the stability region for a stabilized orbit
in the single--scroll chaos is shown. The corresponding Ljapunov 
number $L=-2.15\pm 0.04$ has been calculated from the Poincar\'e
map. The stability diagram includes the stability region
of simple difference control as the special case $(N=0)$. 
Stabilization of the same periodic orbit in the 
double--scroll chaotic attractor is not possible 
with simple difference control.
This is due to a Ljapunov number of $L=-3.27\pm 0.08$ for 
which the method is predicted to fail.

\typeout{Fig.6 goes here!}

\begin{figure}[htbp]
\epsfig{file=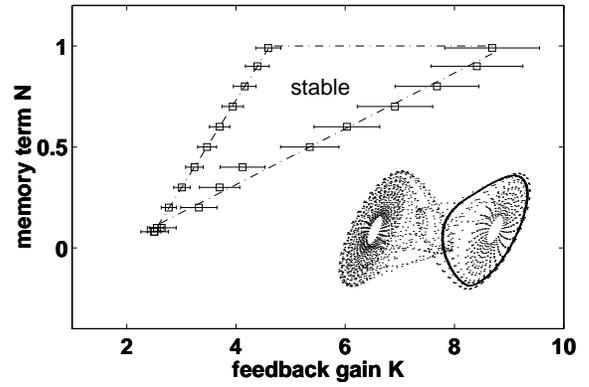, angle=0,width=3in}
\caption{Stabilization of an orbit in the double-scroll chaotic attractor:
In agreement with theory, control without
 memory (N=0) fails
due to a Ljapunov number $L<-3$.
Within the triangle the orbit is stabilized by memory
difference control.
Again,  attractor and orbit are shown
in the inset.
\label{knp1dc}}
\end{figure}

\typeout{Fig.7 goes here!}

The stability region of the stabilized orbit 
in the single-scroll (Fig.~\ref{knp1sc})
and double-scroll attractor (Fig.~\ref{knp1dc})
have a broad overlap. Thus  it is possible to choose
universal values of $(K,N)$ that allow 
tracking of an orbit from one regime to the other
without changing parameters of the controlling circuit.

\typeout{Fig.8 goes here!}

\begin{figure}[htbp]
\epsfig{file=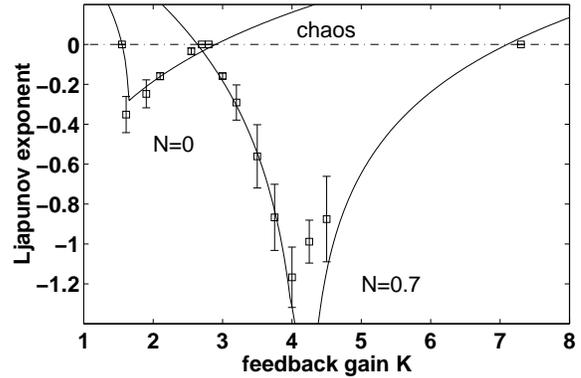, angle=0,width=3in}
\caption{Measured Ljapunov exponents of the control transients for  $N=0$
and
$N=0.7$ (single scroll regime)
compared to 
 theoretical Ljapunov exponents (lines)
 from the eigenvalue analysis.
\label{fig:kl} 
\label{fig:diffkontrljaptheo}}
\end{figure}

The significant improvement of memory difference control compared to simple
difference control is demonstrated  by the estimation of Ljapunov exponents 
(contraction rates) from the transient.
Fig.~8 shows the stability regimes for different feedback gains $K$. 
Simple difference control corresponds to $N=0$, 
and MDC to $N=0.7$. The range of controllability is broadened and the
Ljapunov exponents are smaller, equivalent to faster convergence. 
The measurements are in good
agreement with our theoretical predictions
(Fig.~\ref{fig:diffkontrljaptheo}). 
Due to noise-induced loss of control, it was impossible to obtain
reliable measurements for large $K$ of the case $N=0.7$.

\section{Re-estimation of Ljapunov numbers from the borders of stability}
From the borders and corners of the measured stability region
the exact values of the Ljapunov number 
of the controlled cycle are re-estimated,
similar to the approach used in \cite{schusterniebur}
for OGY control of the H\'enon map.
Since two values of $N$ are exactly one, four of the six coordinate
values can be used to determine $L$ and $M$
by a least-square fit weighted by the variances of the
measured values giving
\begin{eqnarray}
L_{\sf ss} \hspace*{0.15em} = -2.069 \pm  0.03 & ~~~~~~~~ & 
L_{\sf ds} \hspace*{0.2em} = -3.24 \pm 0.03
\nonumber
\\
M_{\sf ss} = \hspace*{0.15em} 0.376 \pm 0.015 & ~~~~~~~~ & 
M_{\sf ds} = \hspace*{0.8em} 0.488 \pm 0.005 
\nonumber
\\
 \label{reest_val1}
\end{eqnarray}
for the orbit stabilized in the single scroll ({\sf{}ss}) and
double scroll ({\sf{}ds}) attractor, respectively.
These values are in good agreement with the values 
given in Section~\ref{chuacontrolsection}
which were obtained from the Poincar\'e{} map (Fig.~\ref{poincarefit}).

\section{Conclusions}

The dynamical behavior and stability conditions of
difference control and memory difference control
are fundamentally different from 
the stability conditions~\cite{just} of
time-continuous Pyragas control~\cite{pyragas92}.
In this paper we introduced and discussed memory difference control 
as a powerful method to stabilize unstable fixed points
even in the presence of parameter drift or delayed measurement.
Investigations of the Chua oscillator circuit 
demonstrated the reliability of the method.

Memory difference control overcomes the Ljapunov number 
limitations of difference control and thus appears to be 
superior both to OGY and Pyragas control schemes.

\begin{acknowledgments}
This work was supported by the Deutsche Forschungsgemeinschaft 
(DFG 185/10, 308/4-1).
The authors want to thank Dr.~Thomas Klinger for inspiring discussions 
and his careful reading of the manuscript. 
\end{acknowledgments}




\begin{thebibliography}{22}
\bibitem{dynamicsreduce}
Once the stable and unstable directions are known, the
Poincar\'e iteration dynamics reduces to the motion in
unstable subspace. Note that the unstable subspace needs only as many 
dimensions as the system has unstable eigenvalues.
In the simplest case, the Ljapunov exponents are of signature
$(+,0,-)$ and the unstable subspace of dimension~1.
\bibitem{remarkmemoryinx}
One could think of extending difference control 
by using delayed differences
$x_{t-i}-x_{t-i-1}$ of the system variable
instead of previous control values $r_{t-i}$.
This is an approach that improves OGY-fashioned control 
if the system is measured with {\sl only one step} 
delay~\cite{claussenthesis}.
Unfortunately, for difference control
one does not obtain a stable control scheme
from this approach.
\bibitem{thirditerate}
Expressing $\vec{x}_{t+1}$ in terms of $\vec{x}_{t-1}$
and $\vec{r}_{t-1}$ shows that $\vec{x}_{t+1}$ vanishes if $K$ and $N$
take the values in eq.~(\ref{eq:verbdiffkont}).
As $\vec{r}_t=-M^{-1} L \vec{x}_{t-1}$, this enforces 
also $\vec{r}_{t+2}$ to vanish. 
Thus, independent of the dimension of $\vec{x}$ and $\vec{r}$,
any initial condition $(\vec{x}_1,\vec{x}_0,\vec{r}_0)$ 
evolves as $(0,\vec{x}_1,\vec{r}_1)$, $(0,0,\vec{r}_2)$ to
$(0,0,0)$ in the {\sl third} iterate. 
Therefore all eigenvalues of the matrix in eq.~(\ref{eq:mdcdynamik})
with parameter choice eq.~(\ref{eq:verbdiffkont})  are zero.
\end{thebibliography}
 \end{document}